\documentclass{kluwer}
\usepackage{graphicx}

\begin{document}
\begin{article}
\begin{opening}
\title{On the outer boundary of the sunspot penumbra}
\author{B\'ela \surname{K\'alm\'an}}
\runningauthor{B. Kalman}
\runningtitle{Penumbra boundary}
\institute{Heliophysical Observatory of the Hungarian Academy of Sciences\\
         P.O.Box 30., H-4010 Debrecen, Hungary. (e-mail: kalman@tigris.klte.hu)}
\date{}

\begin{abstract}
Comparison of photographic observations and vector-magnetograph
measurements demonstrate, that the outer boundary of the sunspot penumbra ---
even in complex sunspot groups --- closely follows the 0.075 T isogauss line
of the total value of the magnetic field, corresponding approximately to the
equipartition value in the photosphere. Radio observations also show this
feature. The thick penumbra model with interchange convection \cite{JS}
gives the best explanation of the penumbral structure.
\end{abstract}

\end{opening}

\section{Introduction}

  The penubra is the least understood structure of sunspots. Although known
from the time of Galileo (\opencite{Galileo}; \opencite{Scheiner}), its radial
filamentary structure was first recognized with the large telescopes of the
$19^{th}$\ century \cite{Secchi}. The first good-quality photographs
of the sub-arcsecond penumbral fibrils were made with Stratoscope I in 1957
\cite{Danielson}. Today speckle-restored observations allow to study
penumbral details with ground-based telescopes too (\opencite{Denker},
\opencite{Sutter}).

Real understanding of the physical processes in sunspots
began in 1908, when Hale observed strong magnetic
fields in them \cite{Hale}. After almost a century, the physics of the sunspot
umbra is more or less clear, but the processes occurring in the penumbra
even in a regular round sunspot are not fully understood yet \cite{TW}.
In the umbra the strong, almost vertical magnetic field suppresses the
convection, thereby reducing the energy reaching the photosphere, so the
temperature and brightness is reduced in these places. Unanswered questions
are in the remaining umbral structure and in the continuation below the
surface. Recently helioseismology gives the possibility to investigate
the subsurface structure and dynamics of sunspots (e.g. \opencite{Koso};
\opencite{Zhao}), but the spatial and temporal resolution of these
measurements is fairly low yet.

The penumbra is entirely different, in it the magnetic field may be even
horizontal (the polarity dividing line between umbrae of different
polarities in sunspot groups often lies in the penumbra),
but in most places the field is inclined considerably to the surface.
As the photospheric plasma must obey the frozen-in condition
of magnetohydrodynamics, the penumbral filaments must be
aligned along the magnetic field, but at the same time the
surface of the penumbra has only an insignificant tilt to the horizontal.
So the penumbral structural elements lie in vertical planes, defined by
the horizontal component of the magnetic field \cite{PU}. The absolute value
of the magnetic field does not change significantly between bright and
dark fibrils, but the inclination to the surface is definitely smaller
in the dark elements (\opencite{Title93}; \opencite{Wiehr00}). Earlier,
low-resolution observations found horizontal field
at the penumbra-photosphere boundary \cite{BS}, but the latest measurements
show, that both the inclination and the magnetic field value is different
from zero at this boundary (\opencite{Skumanich}; \opencite{SRL};
\opencite{Solanki} and references therein; \opencite{Martinez};
\opencite{Wiehr99}).

Recently the fine structure of the penumbral magnetic and velocity
fields was investigated in several papers, using new observational
and reduction methods (\opencite{Stanch}; \opencite{Rudi98},
\citeyear{Rudi99}; \opencite{Martin00}; \opencite{West01a},
\citeyear{West01b}). These are aimed at understanding the
structure and physical processes in the penumbra, but the
more general question, namely what physical quantity
determines the structural boundary between the photosphere
and the penumbra, is also of considerable interest.

The data mentioned above were obtained mostly for regular, round sunspots,
because their structure was supposed symmetrical and so can be described
more easily. However investigation of the interdependence of penumbral
structure and magnetic field is interesting also in complex sunspot
groups for clarifying the processes shaping the penumbra. This paper
describes the behavior of the magnetic field vector on the
penumbra -- photosphere boundary. In Section 2 the observations and
results are presented, in Section 3 these results are discussed in
connection with other data from the literature, finally in
Section 4 the conclusions are drawn.

\section{Observations}

For the study of the magnetic field at the penumbra -- photosphere
boundary four vector magnetograms of the sunspot group
NOAA 6555 were selected, together with photographic observations
of this group. Magnetograms were obtained by the NASA Marshall Space Flight
Center (MSFC) vector-magnetograph \cite{Hagyard}. Photoheliograms were taken
from the Debrecen Heliophysical Observatory archives, they were acquired at
its Gyula Observing Station \cite{Dezso}, and were selected according to their
quality and near-simultaneity with the magnetograms. Figure 1 displays
the photoheliograms for 4 consecutive days in March, 1991,
together with the longitudinal (line-of-sight) magnetic field component and
with maps of the absolute value of the magnetic field. The longitudinal
magnetograms demonstrate, that NOAA 6555 was a magnetically complex spotgroup,
with interesting sunspot proper motions and flare activity \cite{Fontenla},
consisting of an old, multiple-umbra following polarity spot. Newly emerging
umbrae moved around this spot from both sides, causing some large flares in this
interaction \cite{flow}. A color variant of Figure 1, allowing a better
comparison of the photospheric image and the magnetic field strength, and
a movie of the proper motions and spot evolution in this active region
is included in the CD-ROM supplement, or can be downloaded from the URL
http:/\negthinspace /fenyi.sci.klte.hu/$\sim$kalman/penumbra/.

\begin{figure}
%\vspace{110mm}
\centerline{\includegraphics[width=118mm]{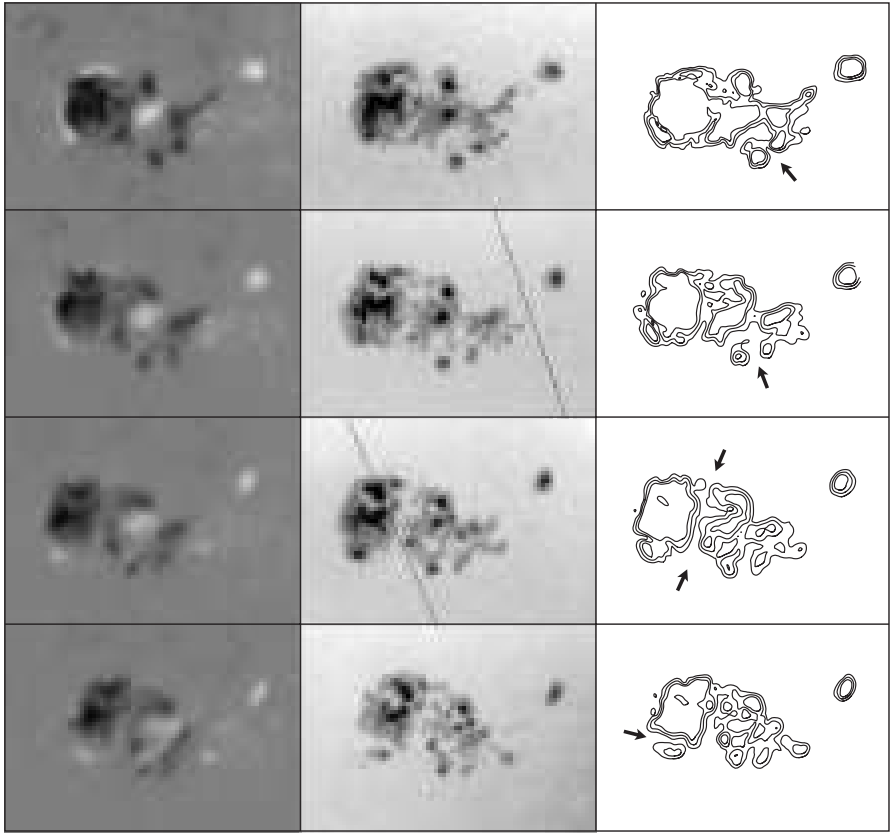}}
  \caption{Longitudinal (line of sight) magnetograms (left column),
           photographs of AR 6555 (middle column), and isogausses of the
           total magnetic field (right column) for 1991 March 23--26.
           Here and on all subsequent images heliographic north is up,
           east to the left.
           On longitudinal magnetograms white is positive (preceding)
           polarity, for the total magnetic field isogausses
           are drawn at 500, 750, and 1000 gauss. The times of observations
           for magnetograms (photographs) from top to bottom are:
           March 23, 18:03 (15:05); March 24, 16:27 (12:33);
           March 25, 13:38 (14:33); March 26, 13:37 (15:54),
           times everywhere in UT.
           (Photographs from Debrecen Heliophysical Observatory,
           magnetograms from NASA Marshall Space Flight Center.) The
           slanted lines on the middle two photographs are the
           images of the (celestial) north-south spider line
           in the photoheliograph.
           The outer border of the penumbra follows the isogauss lines
           of the total magnetic field, and the evolution of both proceeds
           similarly. This is especially well visible at places marked by
           arrows in the right column. A color variant of this figure
           is included in the CD-ROM supplement.}
  \label{mgram}
\end{figure}

On Figure 1 we can see the complex magnetic structure of the active region,
longitudinal magnetic field maps show different polarities in the same penumbra.
On the other hand, on the magnetic field absolute value maps the outer boundary
of the penumbra closely follows the 0.075 T (750 gauss) isoline. This correspondence
is well visible in the process of evolution of the sunspot group, as the penumbra
and the total field maps evolve similarly (see the places marked by arrows on
Figure 1). As the resolution of the magnetograms
is lower, than that of the photographs, the 0.05 -- 0.1 T band
of the absolute value of the magnetic field is indicated on Figure 1.
Large gradients of often opposite polarity
fields paired with the low resolution of the magnetogram can shift the
isogauss line a little, but there is a good overall correspondence between
the outer boundary of the penumbra and the isogausses of the absolute value
of the magnetic field. Especially good example is the p- (white) polarity
umbra, emerging right at the southeast border of the multiple f-polarity
umbra on March 23, and gradually detaching from the common penumbra
to March 26. The isogausses  in the right column of Figure 1 follow
this evolution.

\section{Discussion}

Earlier observations already reported field strength values of about
0.075 -- 0.080 T at the outer boundary of round sunspots
(e.g. \opencite{SRL}; \opencite{Martinez}).
Such an almost constant value of field strength
at the outer boundary of the penumbra can be observed also
for complex sunspot groups, as shown on Figure 1. The observed
constancy of the absolute value of the magnetic field on the
penumbral outer boundary was already mentioned earlier \cite{Krao},
but without interpretation.

The same is true for the recent, high-resolution measurements.
On Figure 3. of \inlinecite{Stanch} the penumbral values of the
magnetic field strength $\vert$ \textit{\textbf{B}} $\vert$ are
all above $\sim$0.08 T. \inlinecite{West01a} find 0.05--0.1 T
on the penumbra-photosphere boundary, increasing with height.
This supports the results of radio observations (see below),
but their analysis can be influenced by the fine structure
of the magnetic and velocity fields \cite{Martin00}.

The value of 0.075 T is approximately equal to the equipartition field
value in the photosphere (\opencite{GW}; \opencite{Wiehr96}),
corresponding to the energy of turbulent convective motions.
The exact value of the equipartition field depends on the depth
and the model of the convective zone, also the calibration of
vector-magnetographs is a difficult problem and model-dependent.
Whether the magnetic field falls abruptly to zero at the boundary or
continues to the nearby photosphere, can not be decided at this
resolution of magnetic measurements, but the change is steep,
and magnetic energy, being proportional to the square of the
field strength, changes even quicker (\opencite{Wiehr96},
\citeyear{Wiehr99}).

\begin{figure}
%\vspace{78mm}
\centerline{\includegraphics[width=118mm]{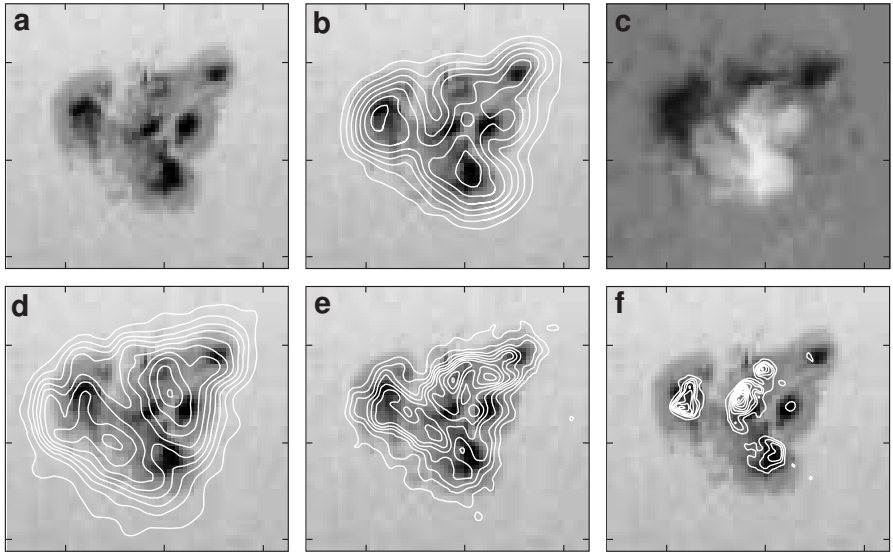}}
  \caption{Optical and radio observations of AR 6615 for
           1991 May 7. (a) photograph of the spotgroup at 06:16 UT
           (Debrecen Observatory), serving also as background for isolines.
           (b) isogausses of the total magnetic field at 17:58 UT
           (NASA MSFC), the lowest one is 450 gauss, every consecutive
           one is 300 gauss higher. (c) Kitt Peak longitudinal magnetogram
           at 14:34 UT, white is positive (preceding) polarity.
           (d)-(f) are VLA observations of gyroresonance emission
           of the solar corona, taken from White et al, (1999). The
           contours begin at 10\% of the maximum brightness and
           are 10\% apart. (d) shows emission at 5 GHz, corresponding
           to a total coronal magnetic field of 450 gauss, (e) and (f)
           display emission at 8.4 GHz and 15 GHz, corresponding to
           750 and 1350 gauss, respectively.}
  \label{radio}
\end{figure}

Another proof of the controlling influence of the total magnetic field value
can be found from radio observations. Gyroresonance emission depends
on the absolute value of the magnetic field in the solar corona above
sunspots. In a series of VLA observations at 1991 May 7 of
NOAA 6615 active region \cite{White}, shown on Figure 2, the 5 GHz
and 8.4 GHz contours, corresponding to 0.045 T and 0.075 T, respectively,
follow nicely the outline of the penumbra of this
complex and compact sunspot group, whereas the 15 GHz emission (0.135 T)
is observed mainly above umbrae (and in the middle of the group above the
neutral line in a $\delta$-configuration, which leads to stronger coronal
activity).

Substantial amount of magnetic flux, and even more heat flux leaves the sunspot
through the penumbra, so it needs lateral energy flow from the surroundings.
This is possible through interchange convection (\opencite{Schmidt};
\opencite{Jahn}; \opencite{JS}), in which magnetic fluxtubes, or sheets
are heated at the outer boundary of the spot (the magnetopause),
then moving upwards and inwards in a vertical plane they supply energy
to the penumbra, whereas cooler material flows downwards and outwards.
Observations support this model: Penumbral filaments lie in vertical planes,
defined by the horizontal component of the magnetic field \cite{PU},
inclination (but not strength) of the field varies in bright and dark
filaments (\opencite{Title93}; \opencite{Wiehr00}). SOHO MDI measurements
\cite{Norton} show enhanced power of intensity oscillations in the range
0.5--1.0 mHz (16.7--33.3 min period) in a ring with filamentary structure
right at the penumbra-photosphere boundary, and magnetic field strength
oscillations also in this period range show filamentary structure in the
penumbra, just as it should be for the interchange convection (Figure 3).

The thick penumbra model (\opencite{JS}, \opencite{Jahn96}) supposes,
that from some depth to the surface the the magnetopause
(penumbra - photosphere boundary) transmits some energy from the
surrounding convective zone, and this energy is distributed
in the penumbra by the interchange convection.
This model uses the monolithic sunspot convention, i.e. the magnetic
field of the sunspot is represented by a single fluxtube of
varying cross-section from the surface to the depth of 15-20 Mm.
The depth of the penumbra is supposed to be about 4-5 Mm. Recent
results \cite{Zhao} seem to contradict the monolithic sunspot model,
showing strong transverse flows at depth $\approx$5Mm.

\begin{figure}
%\vspace{59mm}
\centerline{\includegraphics[width=118mm]{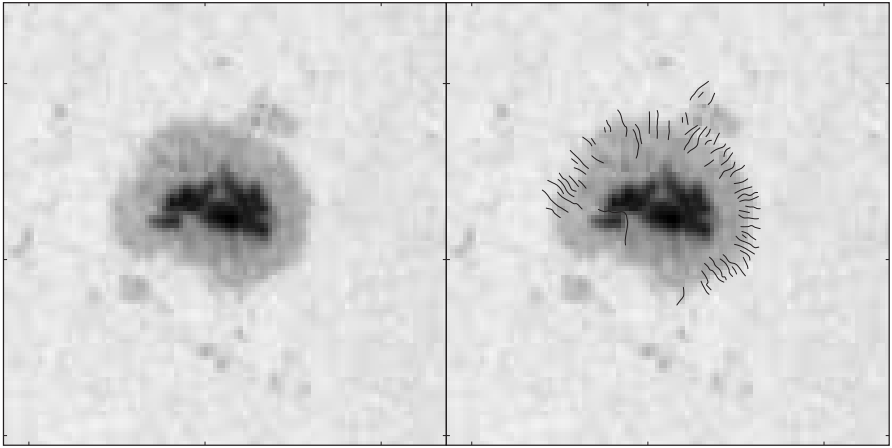}}
  \caption{MDI observations of the leader spot of AR 8113 for
           1997 December 2, taken from Norton et al., (1999). The
           left image is the average photospheric intensity,
           on the right image the places are marked, where
           the intensity fluctuations (grey dots) or the
           longitudinal magnetic field fluctuations (black lines)
           in the 0.5--1.0 mHz range (16.7--33.3 min period range)
           are maximal.}
  \label{MDI}
\end{figure}

It is instructive to follow the change of the equipartition magnetic
field value  with depth, and compare with the magnetic field at the
magnetopause. Figure 4 shows such a comparison, where the
equipartition field values were computed from the energy density
of the convective motions (R. Stein, private communication)
in a 9 Mm deep simulation of the convective zone \cite{Stein},
the magnetic field value at the magnetopause is extrapolated down
from the surface value, scaled according to the the cross-section
of the model of Jahn and Schmidt. The two curves intersect at the
depth of about 6 Mm, which indicates a change of the type
of interaction at this depth. This gives a natural explanation of the
depth of the thick penumbra model: above 6 Mm the energy,
carried by convective motions, can partly penetrate through
the magnetopause. The controversy between the monolithic
and the cluster model \cite{Zhao} can be resolved, if a change
of type with the evolution of the sunspot is supposed
(Kalman, in preparation): younger sunspots at emergence have deep
connections, which is severed later by the turbulent convection
below 6 Mm. Stable and decaying sunspots are shallow, and
are hold together by their moat cell (\opencite{Hurl};
\opencite{Zhao}) until the convection finally erodes their
magnetic field concentration.

\begin{figure}
%\vspace{90mm}
\centerline{\includegraphics[width=118mm]{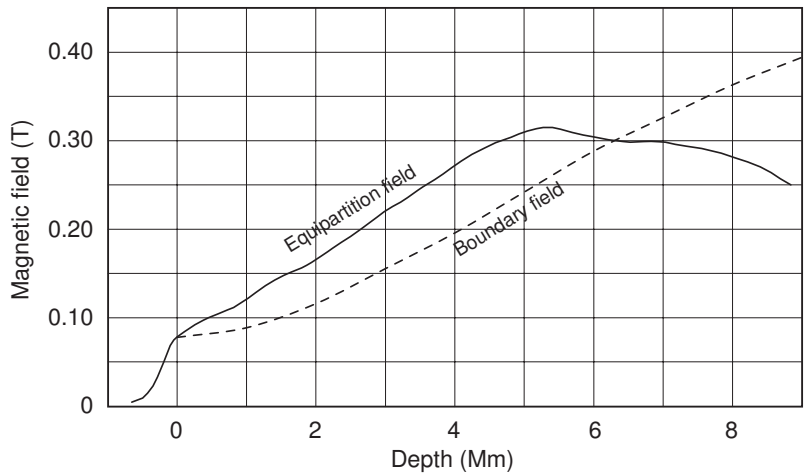}}
  \caption{The variation with depth of the equipartition magnetic field
           (Stein, private communication) in a
           9 Mm deep simulation of the convective zone (Stein and Nordlund,
           1994) and of the field strength on the outer boundary of the
           sunspot (the magnetopause) according to the model of Jahn and
           Schmidt (1994).}
  \label{Equi}
\end{figure}

\section{Conclusion}

Observations show, that the outer boundary of the sunspot penumbra
even in complex active regions follows the isogauss line of the
total magnetic field corresponding approximately to the
equipartition field value at the surface (Figure 1).
The observations also support the thick penumbra model with
interchange convection \cite{JS}, the change of the ratio
of the equipartition field  value and the field value at the
magnetopause boundary with depth gives a natural depth of
the penumbra about 5--6 Mm. The magnetic field in the
penumbra is not strong enough to stop the convection,
but severely alters its nature,
leading to formation of interchange convection. The enigmatic penumbra
at last seems to be understandable.

\acknowledgements
The author thanks Dr. Mona Hagyard of NASA MSFC for sending the magnetograms
used, Drs. R.F. Stein and H. Spruit for sending their respective convective
zone model data, and Drs. H.U. Schmidt and K. Jahn for informations about their
thick penumbra model, also the anonymous referee for suggesting more
literature supporting the conclusion.
Kitt Peak Observatory magnetogram was taken from
NSO/Kitt Peak Internet archives. NSO/Kitt Peak data used here are produced
cooperatively by NSF/NOAO, NASA/GSFC and NOAA/SEL.
Debrecen Observatory photographic observations were made by I. Lengyel.
This research has made use of NASA's Astrophysics Data System
Abstract Service. Part of this work was supported by grant T-025737
of the Hungarian Scientific Research Fund (OTKA).

\end{article}

\end{document}